\def\astroph{1}

\ifnum\astroph=0
\documentclass[12pt,preprint,tighten]{aastex}
\else
\documentclass{emulateapj}
\usepackage{apjfonts}
\fi

\usepackage{natbib}
\usepackage{amssymb}
\usepackage{bm}
\usepackage{epsfig}
\usepackage{citesort}

\newcommand{\mpch}{\textrm{\,Mpc h}^{-1}}
\newcommand{\kms}{\textrm{\,km s}^{-1}}
\newcommand{\Eq}[1]{Eq.\ (\ref{#1})}

\newcommand{\WE}[3]{\rule[-2ex]{0pt}{5ex}${#1\,\,}^{\textstyle{#2}}_{\textstyle{#3}}$}

\begin{document}
\title{The velocity field of the local universe from measurements of type Ia
supernovae}

\author{Troels Haugb{\o}lle\altaffilmark{1},
Steen Hannestad\altaffilmark{1}, Bjarne Thomsen\altaffilmark{1},
Johan Fynbo\altaffilmark{2}, Jesper Sollerman\altaffilmark{2},
Saurabh Jha\altaffilmark{3}
} \altaffiltext{1}{Department of Physics and Astronomy, University
of Aarhus, DK-8000 Aarhus C, Denmark} \altaffiltext{2}{DARK
Cosmology Centre, Niels Bohr Institute, University of Copenhagen,
Juliane Maries Vej 30, DK-2100 Copenhagen \O, Denmark}\altaffiltext{3}{Kavli
Institute for Particle Astrophysics and Cosmology, Stanford Linear
Accelerator Center, 2575 Sand Hill Road MS 29, Menlo Park, CA 94025}
\email{haugboel@phys.au.dk}

\begin{abstract}
We present a measurement of the velocity flow of the local
universe relative to the CMB rest frame, based on the Jha, Riess
\& Kirshner (2007) sample of 133 low redshift type Ia supernovae.
At a depth of 4500 $\kms$ we find a dipole amplitude of $279 \pm
68 \kms$ in the direction $l = 285^\circ \pm 18^\circ$,
$b=-10^\circ \pm 15^\circ$, consistent with earlier measurements
and with the assumption that the local velocity field is dominated
by the Great Attractor region. At a larger depth of 5900 $\kms$ we
find a shift in the dipole direction towards the Shapley
concentration. We also present the first measurement of the
quadrupole term in the local velocity flow at these depths. Finally, we
have performed detailed studies based on N-body simulations of the
expected precision with which the lowest multipoles in the
velocity field can be measured out to redshifts of order 0.1. Our
mock catalogues are in good agreement with current observations,
and demonstrate that our results are robust with respect to
assumptions about the influence of local environment on the type
Ia supernova rate.
\end{abstract}

\section{Introduction}

Distant type Ia supernovae (SNe Ia) have been used to probe the expansion
rate of the Universe out to redshifts of order 1.5
\citep[see e.g.][]{riess:2004}. These
measurements were crucial in establishing the current standard
model of cosmology in which roughly 30\% of the energy density is
in the form of non-relativistic matter, whereas roughly 70\% is in
the form of a dark energy, a component with negative pressure.

The power of supernova surveys as cosmological probes depends
on a precise measurement of the luminosity distance to the individual
supernovae - but the supernova host galaxies do not follow the Hubble flow.
They have peculiar velocities, induced by the underlying gravitational
potential, and the measured luminosity distances and redshifts are
perturbed in correlation with the large scale structure
\citep[see e.g.][]{miller:1992}.

The measured fluctuations in the luminosity distances from nearby
supernovae surveys can be related to the local variation in the
Hubble parameter \citep{Riess:1995}, and the local large scale
structure, and can be quantified in terms of their correlation
functions. This possibility has been studied in several recent
papers \citep{Bonvin:2005ps,Bonvin:2006en,Hui:2005nm} \citep[see
also][]{sugiura:1999} using analytical methods.
\citet{Bonvin:2006en} detected the dipole term with respect to the
CMB at roughly 2$\sigma$ confidence. The variation of the monopole
contribution with redshift was studied by \citet{Zehavi:1998gz}
using the same supernova sample as \citep{Riess:1995}.

In contrast to measurements of the density field, velocity field
measurements are much less sensitive to selection bias, since the
field is measured directly instead of summed up from number counting,
but it is very sensitive to uncertainties in the measured luminosity distance.
Because of this type Ia supernovae are particularly useful as
probes of the velocity field because of the very small inherent
uncertainty in their luminosity distance. Compared with
measurements using galaxies as standard candles much fewer
supernovae are needed to get a reliable estimate of at least the
lowest multipoles in the velocity field.

The purpose of the present paper is twofold. The first is to
perform a theoretical study of the problem of extracting the velocity
field from supernova data. The second is to use the developed
formalism on the best available data set to extract precise
values of the dipole and quadrupole terms in the local velocity
field.

We first use large scale dark matter N-body models to predict the
observed angular power spectrum of the peculiar radial velocity
field as a function of redshift and to explore its utility to
probe the local velocity field. By using this method we are able
to quantify the effect of various error sources as well as cosmic
variance.

We also use the \citet*{jrk07} (hereafter JRK) sample of nearby
type Ia supernovae to calculate the lowest multipoles of the local
velocity field. We find that both the dipole and the quadrupole
terms are well measured by this sample. Higher multipoles cannot
be reliably estimated with the JRK sample because of sparse
sampling. In \citet{jrk07} this sample was also used to probe the
monopole term in an analysis similar to that of
\citet{Zehavi:1998gz}. Similar evidence of a local void was found
in \citet*{jrk07}, even taking a data set which is completely
disjoint from that in \citet{Zehavi:1998gz}. In the following we
will concentrate on the dipole and quadrupole terms in the JRK
sample since the monopole has already been exhaustively discussed
in \citet{jrk07}.

In Section 2 we discuss the formalism used to derive the velocity
field from magnitude measurements. In Section 3 we present the
analysis tools used to extract the multipole components of the
velocity field from a finite sample of supernovae. In Section 4 we
use catalogues from N-body simulations to study the expected
properties of the velocity field, including the precision with
which it can be probed by supernova surveys. We apply the same
formalism to the JRK sample and provide a detailed discussion of
the results in Section 5. Finally, Section 6 contains a comparison
with other measurements of the local velocity field, and in
Section 7 we provide our conclusions.

In the following we will take $c=1$.

\section{Distance measures and radial velocities}

The luminosity distance, $d_L$, to a supernova at redshift $z$ is
given by the definition
\begin{equation}
d_L = \sqrt{\frac{L}{4\pi F}}\,,
\end{equation}
where $F$ is the observed flux and $L$ the luminosity,
or equivalently by
\begin{equation}
m = 5 \log_{10} \left(\frac{d_L}{1 \, {\rm Mpc}}\right) + M + 25\,,
\end{equation}
where $m$ is the apparent and $M$ the absolute magnitude. The
angular-diameter distance, $d_A$, to the same supernova is defined
as
\begin{equation}
d_A = \frac{d_L}{(1+z)^2}
\end{equation}

In a homogeneous and isotropic universe the two distance
measures, $d_L$ and $d_A$, are given by
\begin{equation}\label{eq:dA1}
d_A(1+z) = \frac{d_L}{(1+z)} = \frac{1}{H_0} \cases{\frac{1}{\sqrt{\Omega-1}}
\sin\left(\sqrt{\Omega-1}\,\, I\right) & $\Omega > 1$ \cr I & $\Omega=1$
\cr \frac{1}{\sqrt{1-\Omega}} \sinh \left(\sqrt{1-\Omega}\,\, I\right)
& $\Omega < 1$ \cr}\,,
\end{equation}
where
\begin{equation}
I = H_0 \int^{z}_0 \frac{dz}{H(z)} = \int^{1+z}_1
    \frac{dx}{\sqrt{\Omega_{m}x^3 + \Omega_{K}x^2 + \Omega_{\Lambda}}}\,,
\end{equation}
\vskip 0.2cm
By taking the logarithmic derivative of \Eq{eq:dA1} with
respect to $(1+z)$ in the special case of a flat universe ($\Omega = 1$) we obtain
\begin{equation}\label{eq:alpha}
\alpha_L(z)-1=\alpha_A(z)+1=\frac{1}{H(z)d_A(z)}=\frac{(1+z)^2}{H(z)d_L(z)}\,,
\end{equation}
where
\begin{eqnarray}
\alpha_L = \frac{d \ln d_L}{d \ln (1+z)}\hskip 1cm{\rm and}\hskip 1cm
\alpha_A = \frac{d \ln d_A}{d \ln (1+z)}
\end{eqnarray}

However, in a perturbed universe, the luminosity distance depends
on the detailed trajectory of the individual photons from the
supernova to the observer. At moderately high redshifts ($z
\gtrsim 0.5$) the contribution arising from lensing by intervening
matter dominates, while at low redshifts ($z \lesssim 0.5$) the
contribution from the peculiar velocities of the supernova host
galaxy relative to the observer dominates. In this paper we are
only concerned with the local universe. From now on we will
therefore only consider the contribution from the  peculiar
velocities in determining distances. For a given supernova and
observer, each with some peculiar velocity, the measured redshift
$z$, angular-diameter distance, $d_A$, and luminosity distance,
$d_L$, are modified according to \citep{Hui:2005nm,Bonvin:2006en}
\begin{eqnarray}
1+z & = & (1-v_0\cdot n)(1+\bar{z})(1+v_r) \\
d_A & = & \bar{d}_A(\bar{z})(1+v_0\cdot n) \\
d_L & = & \bar{d}_L(\bar{z})(1+v_0\cdot n)(1+v_r-v_0\cdot n)^2\,,
\end{eqnarray}
where by definition
\begin{eqnarray}
d_L & \equiv & d_A(1+z)^2 \\
\bar{d}_L(\bar{z}) & \equiv & \bar{d}_A(\bar{z})(1+\bar{z})^2
\end{eqnarray}
A bar $\bar{\ldots}$ indicates the quantities as measured in a
homogeneous and isotropic cosmology, and $v_r = v_e\cdot n$ is the
velocity of the supernova projected along the direction
from the observer to the supernova. Having measured the redshift, $z$, and
the flux, $F$, of the supernova we can calculate the luminosity
distance, $d_L$, and the angular-diameter distance, $d_A$, to
within a scatter determined by the cosmic variance on the
luminosity, $L$, of a supernova. As we know the peculiar velocity,
$v_0$, of the observer with respect to the CMB with great accuracy
it is useful to collect all known quantities on the left hand side
of the three equations:
\begin{eqnarray}
\label{eq:Z}
1+\tilde{z} & = & (1+z)(1+v_0\cdot n) = (1+\bar{z})(1+v_r) \\
\label{eq:A}
\tilde{d}_A & = & d_A(1-v_0\cdot n) = \bar{d}_A(\bar{z}) \\
\label{eq:L}
\tilde{d}_L & = & d_L(1+v_0\cdot n) = \bar{d}_L(\bar{z})(1+2\, v_r)
\end{eqnarray}
The three quantities $\tilde{z}$, $\tilde{d}_A$, and $\tilde{d}_L$
are the redshift, angular-diameter distance, and luminosity distance
as measured and calculated by an observer at rest with respect to the CMB.
In the following we assume that the measured quantities all have been
corrected to a frame at rest with respect to the CMB. The measured
redshift, $\tilde{z}$, is most transparently split into the cosmological
redshift, $\bar{z}$, and the radial velocity of the supernova, $v_r$, by
inverting \Eq{eq:A} to find $\bar{z}$, which is then inserted into this
alternative form of \Eq{eq:Z}:
\begin{equation}\label{eq:Vr}
v_r = \ln(1+\tilde{z})-\ln(1+\bar{z})
\end{equation}

It is also possible to begin with \Eq{eq:Vr} to obtain the cosmological
redshift, $\bar{z}$, in terms of the measured redshift, $\tilde{z}$,
and the radial velocity, $v_r\,$, of the supernova:
\begin{equation}
\ln(1+\bar{z}) = \ln(1+\tilde{z})-v_r
\end{equation}
We now expand the natural logarithm of the angular-diameter distance,
$\ln\bar{d}_A(\bar{z})$, to first order around the measured redshift,
$\tilde{z}$, and by using \Eq{eq:A} we obtain
\begin{equation}
\ln\tilde{d}_A = \ln\bar{d}_A(\bar{z}) =
                 \ln\bar{d}_A(\tilde{z})-\alpha_A(\tilde{z})v_r\,,
\end{equation}
where $\alpha_A(z)$ for a flat universe is given by \Eq{eq:alpha}.
This equation applies equally well for the luminosity distance,
so we can give a common formula for both distance measures:
\begin{equation}
\ln\left(\frac{\tilde{d}_L}{\bar{d}_L(\tilde{z})}\right) =
\ln\left(\frac{\tilde{d}_A}{\bar{d}_A(\tilde{z})}\right) =
-\alpha_A(\tilde{z})v_r
\end{equation}
This expression can easily be transformed into an equation between
the measured apparent magnitude, $\tilde{m}$, the calculated
apparent magnitude, $\bar{m}(\tilde{z})$, at the measured redshift,
$\tilde{z}$, and the peculiar radial velocity, $v_r\,$, of the
supernova:
\begin{equation}
\tilde{m}-\bar{m}(\tilde{z}) =
5\log_{10}\left(\frac{\tilde{d}_L}{\bar{d}_L(\tilde{z})}\right) =
-(5/\ln 10)\alpha_A(\tilde{z})v_r
\end{equation}
Inverting this equation for the special case of a flat universe
we obtain
\begin{equation}\label{eq:vr1}
v_r = -(\ln 10/5)\left(\frac{H(\tilde{z})\bar{d}_A(\tilde{z})}
                      {1-H(\tilde{z})\bar{d}_A(\tilde{z})}\right)
                      \left(\tilde{m}-\bar{m}(\tilde{z})\right)
\end{equation}
For low redshift supernovae the Hubble parameter, the
angular-diameter distance, and the luminosity distance can all
be expanded in terms of the deceleration parameter, $q_0$\,:
\begin{eqnarray}
H(z)&=&H_0\left(1 + (1+q_0)z\right) \\
\bar{d}_A(z)&=&\frac{z}{H_0}\left(1 - (3+q_0)z/2\right) \\
\bar{d}_L(z)&=&\frac{z}{H_0}\left(1 + (1-q_0)z/2\right)
\end{eqnarray}
When we insert these values into \Eq{eq:vr1} we obtain the radial
velocity of a low redshift supernova
\begin{equation} \label{eq:approxvr}
v_r = -(\ln 10/5)\tilde{z}\left(1+(1+q_0)\tilde{z}/2\right)
                          \left(\tilde{m}-\bar{m}(\tilde{z})\right)
\end{equation}
In this approximation the distance modulus, $m - M$, is given by
\begin{eqnarray}\label{eq:vr2}\nonumber
\bar{m}(\tilde{z}) - M &=&42.3841 - 5 \log_{10} h  \\
                                     && +\, 5\log_{10} \tilde{z}+\,(2.5/\ln 10)(1 - q_0)\,\tilde{z}\,,
\end{eqnarray}
where the Hubble constant as usual is given as
$H_0 = 100 h\kms{\rm Mpc}^{-1}$. The distance
modulus defined in \Eq{eq:vr2} should be compared with the
distance modulus, $\tilde{m}-M$, as measured in a frame at rest
with respect to the CMB. The difference between the apparent
magnitude $\tilde{m}$ and the one measured by the observer, $m$,
is according to \Eq{eq:L},
\begin{equation}
\tilde{m} - m = (5/\ln 10) v_0\cdot n = 2.17 v_0\cdot n\,.
\end{equation}
The approximate equation (\ref{eq:approxvr}) is sufficient for the low redshifts
that we are considering, while at higher redshifts one has to use the correct
form Eq.~(\ref{eq:vr1}), and also take into account other contributions to
$\tilde m - \bar m(\tilde z)$, such as lensing.

\section{Analysis using an angular expansion of the radial velocity field}\label{sec:analysis}

We analyse both the mock catalogues discussed in the next section
and the real data set (see section \ref{sec:data}) using the same
technique. In practice we decompose the field into spherical
harmonics.

The radial velocity is a real scalar field, and on a spherical
shell of a given redshift it can be decomposed into spherical
harmonics
\begin{eqnarray}
v_r &=& \sum_{l = 0}^\infty \sum_{m=-l}^{l} a_{lm}Y_{lm} \nonumber \\
    &=& \sum_{l = 0}^\infty \left\lbrace\sum_{m=1}^{l}(a_{l,-m}Y_{l,-m}+a_{lm}Y_{lm})
      +a_{l0}Y_{l0}\right\rbrace
\end{eqnarray}
Using $a_{l,-m}=(-1)^{m}a_{lm}^{*}$ for the expansion of a real
function and $Y_{l,-m}=(-1)^{m}Y_{lm}^{*}$ we obtain
\begin{eqnarray}
v_r&=&\sum_{l = 0}^\infty\left\lbrace
\sum_{m=1}^{l}\left[2\Re(a_{lm}Y_{lm})\right]+a_{l0}Y_{l0}\right\rbrace
         \\ \nonumber
    &=&\sum_{l = 0}^\infty\left\lbrace \sum_{m=1}^{l}\left[2\Re(a_{lm})\Re(Y_{lm})-
                                           2\Im(a_{lm})\Im(Y_{lm})\right]
          +a_{l0}Y_{l0}\right\rbrace
\end{eqnarray}
However, this applies strictly only if the field can be measured
on the entire sphere. In our case the radial velocity field is
measured for a finite number of directions, so we can only hope to
determine a finite number of $a_{lm}$ coefficients by fitting a
truncated multipole expansion by the method of weighted linear
least squares using $[Y_{l0}, \lbrace 2\Re(Y_{lm}), -2
\Im(Y_{lm})\rbrace , m=1,\ldots,l]$ as basis functions.
Specifically, we solved the problem by a singular value
decomposition.

We follow the procedure by \citet*{copi06} and represent the
$l$'th multipole in terms of a scalar, $A^{(l)}$ and $l$ unit
vectors, $\lbrace \hat{v}^{(l,m)}, m=1,\ldots,l \rbrace$ :
\begin{equation}
f_l(\theta,\phi) =
A^{(l)}\lbrace\prod_{m=1}^{l}(\hat{v}^{(l,m)}\cdot \hat{e})
 - \mathcal{T}_l\rbrace ,
\end{equation}
where $\hat{e} =
(\sin\theta\cos\phi,\sin\theta\sin\phi,\cos\theta)$, and
$\mathcal{T}_l$ is the sum of all possible traces of the first
term. In this representation the multipole expansion up to and
including the quadrupole term takes the following form:
\begin{eqnarray}
v_r(\hat{e}) & = & A^{(0)} + A^{(1)}(\hat{v}^{(1,1)}\cdot \hat{e}) \nonumber \\
             & + & A^{(2)}\lbrace(\hat{v}^{(2,1)}\cdot \hat{e})(\hat{v}^{(2,2)}\cdot \hat{e}) -
(1/3)(\hat{v}^{(2,1)}\cdot \hat{v}^{(2,2)})\rbrace
\end{eqnarray}
Note that $\hat{v}^{(2,1)}$ and $\hat{v}^{(2,2)}$ are ``headless''
vectors only defining a line, not a direction. Equivalently they
define a plane, but they do not define a rotation in
that plane, so the normal to the plane is also headless. By
convention we choose as the first vector, $\hat{e}_1$, the one
with the largest absolute z-coordinate. We can choose $\hat{e}_1$
to point to the hemisphere near the pole without introducing a
negative amplitude $A^{(2)}$, if both $\hat{e}_1$ and $\hat{e}_2$
have their sign changed. Finally we define the normal to the plane
spanned by the two vectors as $\hat{e}_1\times\hat{e}_2$. This is
the {\it polar quadrupole vector}.

From the $a_{lm}$ coefficients the monopole amplitude can be found
as
\begin{equation}
A^{(0)} = \frac{a_{00}}{\sqrt{4\pi}}
\end{equation}
and the dipole amplitude and direction can be found as
\begin{eqnarray}
A^{(1)} & = & (a_{10}^2 + 2{|a_{11}|}^2)^{1/2}
\\ \theta & = & -\tan^{-1}\left(\frac{\Im(a_{11})}{\Re(a_{11})}\right)
\\ \phi & = & \cos^{-1} \left(\frac{a_{10}}{A^{(1)}}\right)
\end{eqnarray}
This is the direction of the maximum of the dipole.
All the higher order multipole vectors are found by using the
program mpd\_decomp by \citet{copi06}.

For the lowest $l$-values the amplitudes in the multipole vector
expansion are related to the usual power $C_l$ as \citep{copi06}
\begin{eqnarray}
C_0&=&4\pi \left[ A^{(0)} \right]^2 \\
C_1&=&\frac{4\pi}{9} \left[ A^{(1)} \right]^2 \\
C_2&=&\frac{4\pi}{75} \left[ A^{(2)} \right]^2\left[ 1 + \frac{1}{3}\left(\hat{v}^{(2,1)}\cdot\hat{v}^{(2,2)}\right)\right] \\
\end{eqnarray}

It should be noted that in general the individual multipole
coefficients obtained in the fit to data can be strongly dependent
on the number of modes included. The reason for this is that the
window function does not cover the entire sky, rather there are
patches with zero coverage. This means that the spherical
harmonics are no longer orthogonal, and shows up as a leakage of
power between different $l$. In fact, this is predicted to be a
significant problem for any harmonic analysis with limited
sampling because the higher order multipoles do contribute
significantly to the rms velocity. In Section \ref{sec:window} we
discuss the implications of sampling for the JRK sample.

\section{Synthetic supernova surveys from Monte Carlo simulations}

\begin{figure}
\begin{center}
\includegraphics[width=0.4 \textwidth]{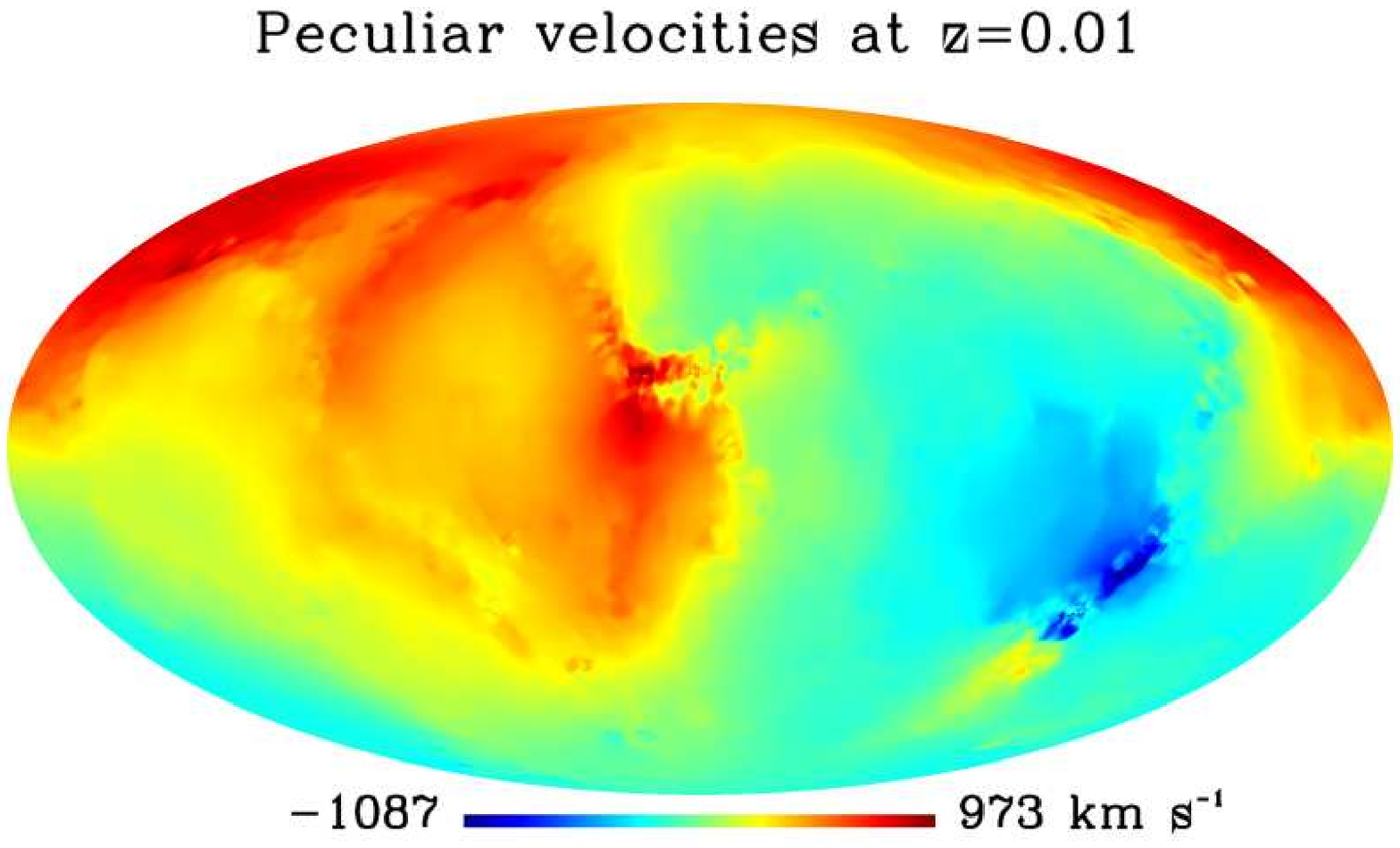}
\includegraphics[width=0.4 \textwidth]{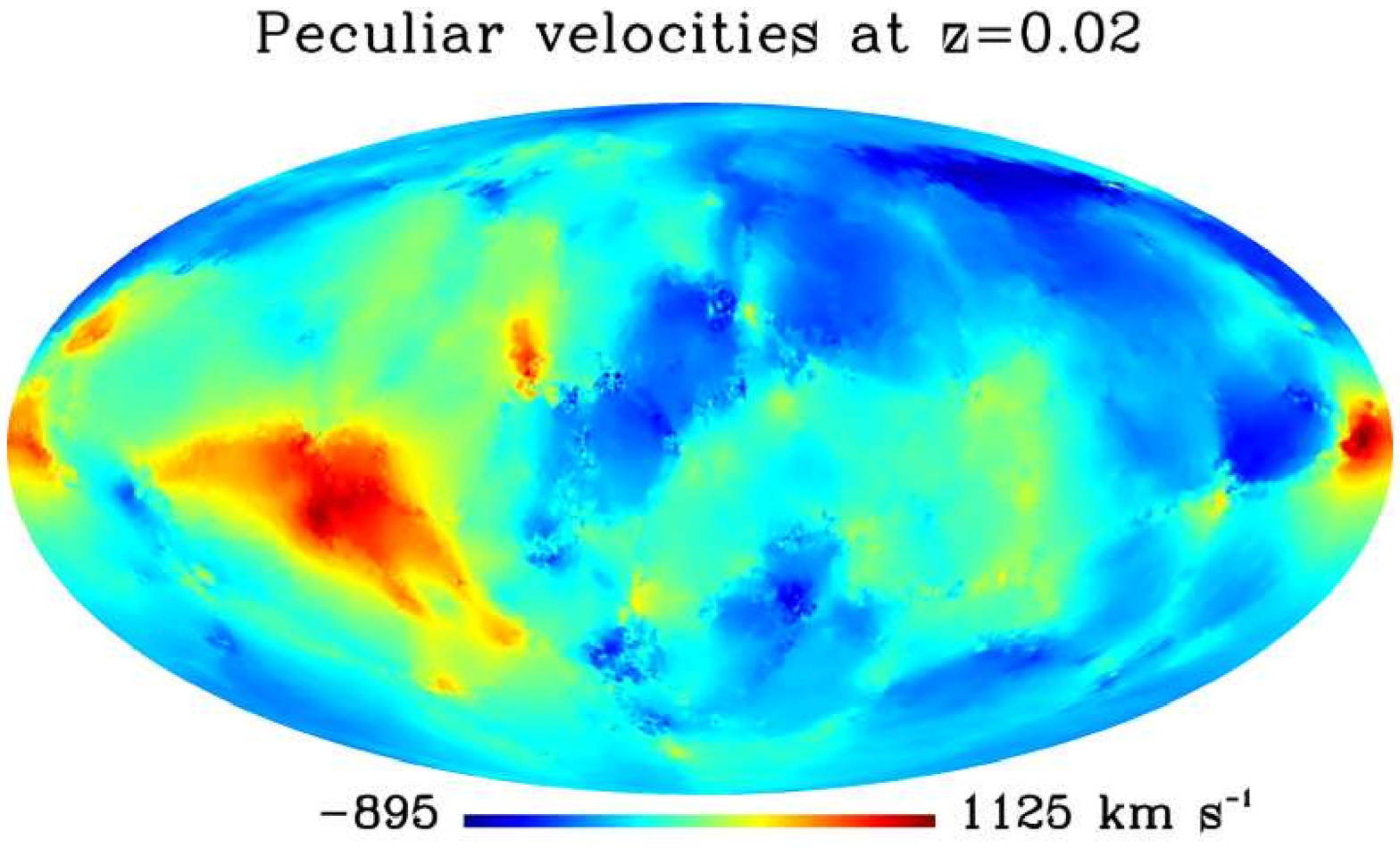}
\includegraphics[width=0.4 \textwidth]{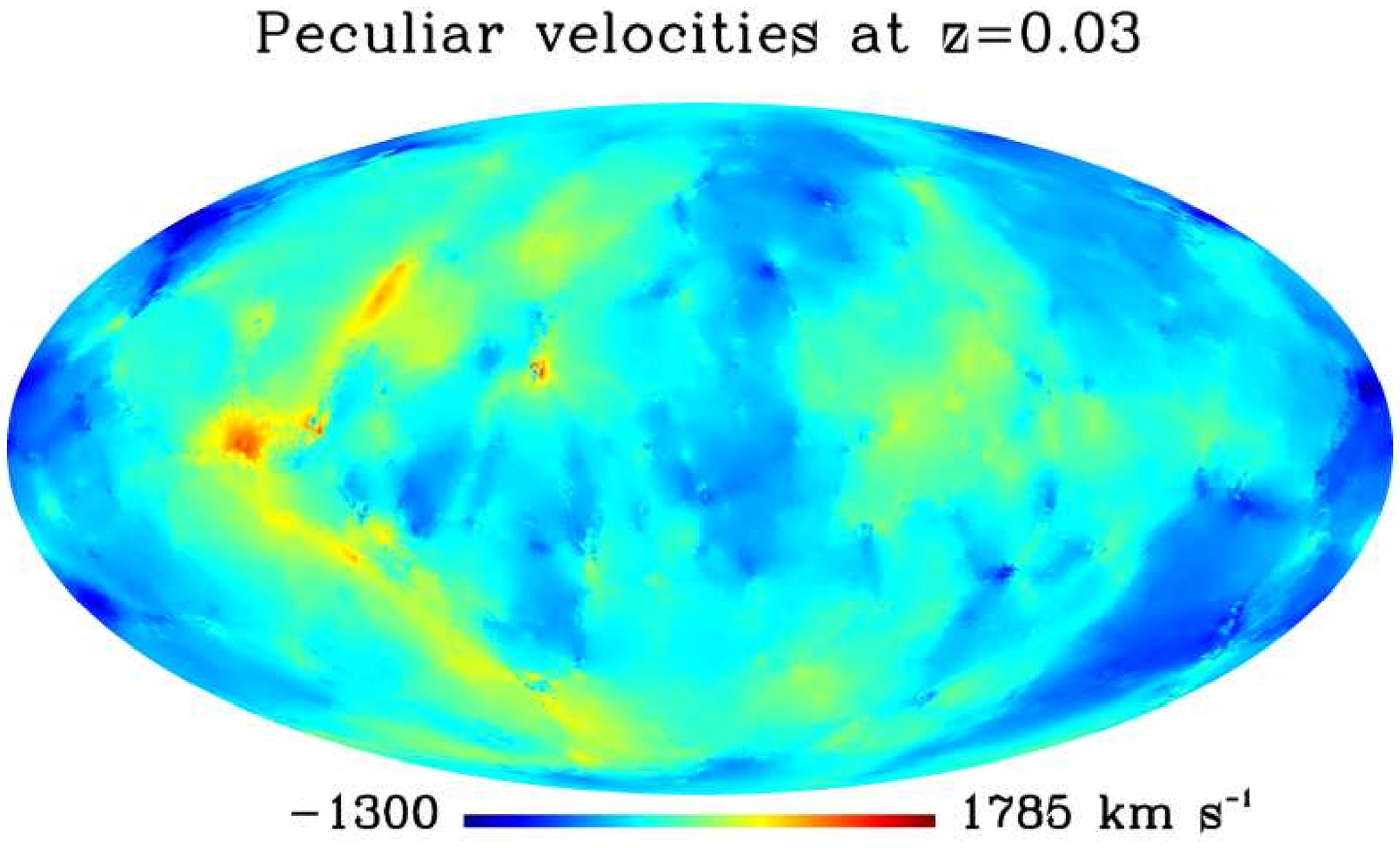}
\includegraphics[width=0.4 \textwidth]{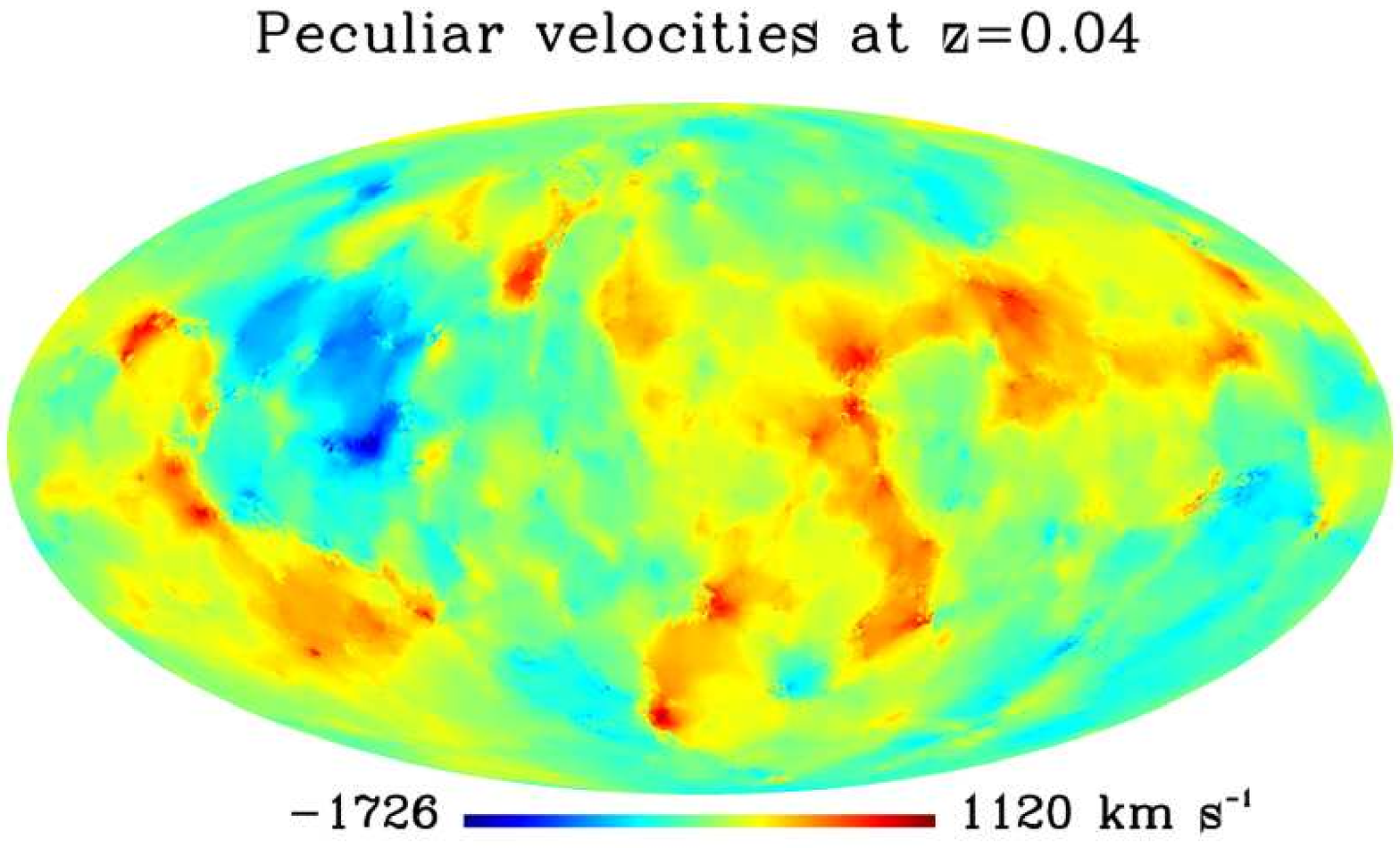}
\caption{The variation in the peculiar velocity at $z=0.01-0.04$.}\label{fig:mollview}
\end{center}
\end{figure}

Before analysing existing data we make mock catalogues of
supernova data based on dark matter N-body simulations. This is done in order
to get an estimate of the various sources of error in such
measurements. The N-body simulations were done using the Gadget-2
code \citep{Springel:2005mi,Springel:2000yr} with a box size of
$800 \mpch$ and $512^3$ \& $768^3$ particles respectively to make
synthetic realisations. The box size is chosen large enough that
the periodic nature of the box does not impact the simulation at
the scales ($z\lesssim0.1$, or $\lesssim 300\mpch$) we are
interested in, and the high resolution run is made to assure that
our results are not dependent on the numerical resolution. In
Fig.~\ref{fig:mollview} is shown a typical all sky map in
Mollweide projection of the peculiar velocity field at redshifts
$z=0.01-0.04$, computed in the CMB rest frame.

\begin{figure}
\begin{center}
\includegraphics[width=0.4 \textwidth]{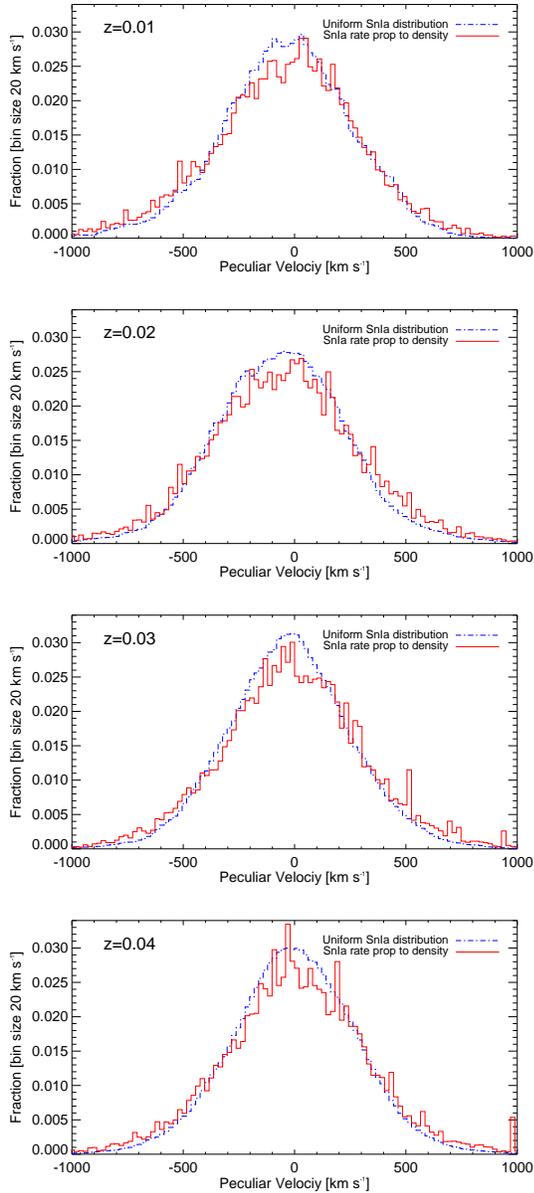}
\caption{The peculiar velocity distribution of supernovae for
different types of scenarios, and at different redshifts.
Sampling with a SNe Ia rate proportional to the density compared
to uniformly on the sky only bias the velocity distribution a
few percent.}\label{fig:sniahisto}
\end{center}
\end{figure}

The formation rate of SNe Ia as a function of the
environment is not well known, though there are indications that
at low redshifts the rate is directly proportional to the stellar mass,
and insensitive to the metalicity \citep{Neill:2006} \citep[though see][and references
therein for indications of a bimodal distribution]{sharon,sullivan}.  On the other hand the
semi-analytic estimates
in the literature \citep[see e.g.][]{Bonvin:2006en,Hui:2005nm} assume a rate which is uniform on the
sky. This is clearly not realistic, but to test the effect on the
luminosity distance distributions we have made synthetic data sets
using both a rate proportional to the mass and an uniform
distribution. In Fig.~\ref{fig:sniahisto} we show
the distribution of peculiar radial velocities in the two models.
The differences between the two scenarios are at a percent level.
Looking at Fig.~\ref{fig:mollview} we see that the radial velocity field is
smooth across voids in contrast to e.g.~scalar fields like the density field.
This is because it is only a pseudo-scalar, and the underlying vector field
can be transported efficiently (i.e.~it is easier to change the direction of a vector,
than transport a scalar quantity).

\subsection{Making a mock supernova survey}

Real measurements rely on a tracer of the matter distribution,
whether it is galaxies or SN Ia. In any case, only a
finite number of objects will be available. Furthermore there will
be a selection bias coming from the presence of galactic
foregrounds etc.
In this work we generate a mock supernova survey using the
following strategy:

{\it a)} The total number of measured SNe in a redshift bin, $N$, is chosen.

{\it b)} Each SN is generated by sampling one of the two probability distributions
described in the previous section.

{\it c)} To each SN a noise component is added, resulting from scatter in
the (stretch corrected) intrinsic luminosity, uncertainty in extinction correction,
measurement errors etc. We describe the errors as a Gaussian error with a
spread of $\Delta m$ on the measured apparent magnitude of the SN.

From this data set the angular power spectrum is calculated. For
each type of simulated survey this task is performed 500 times for
27 different observers, to find the mean and variance of the angular
power spectrum.

\begin{figure}
\begin{center}
\includegraphics[width=0.45 \textwidth]{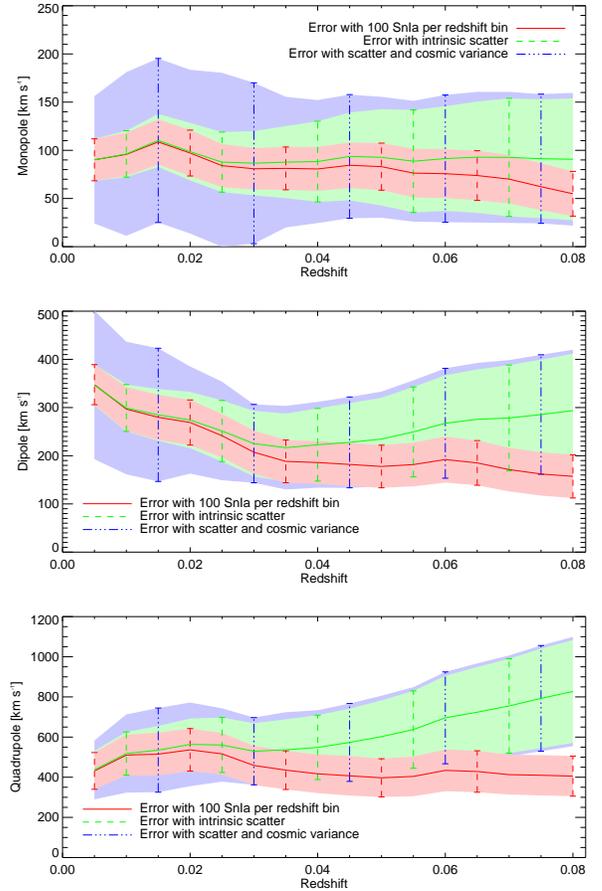}
\caption{Uniform distribution of Supernovae: The amplitudes of the
multipole vectors and the different errors in a SN Ia survey with
100 SNe per redshift bin and an intrinsic scatter in the magnitude
of $\Delta m=0.08$.}\label{fig:multipoles}
\end{center}
\end{figure}

\begin{figure}
\begin{center}
\includegraphics[width=0.45 \textwidth]{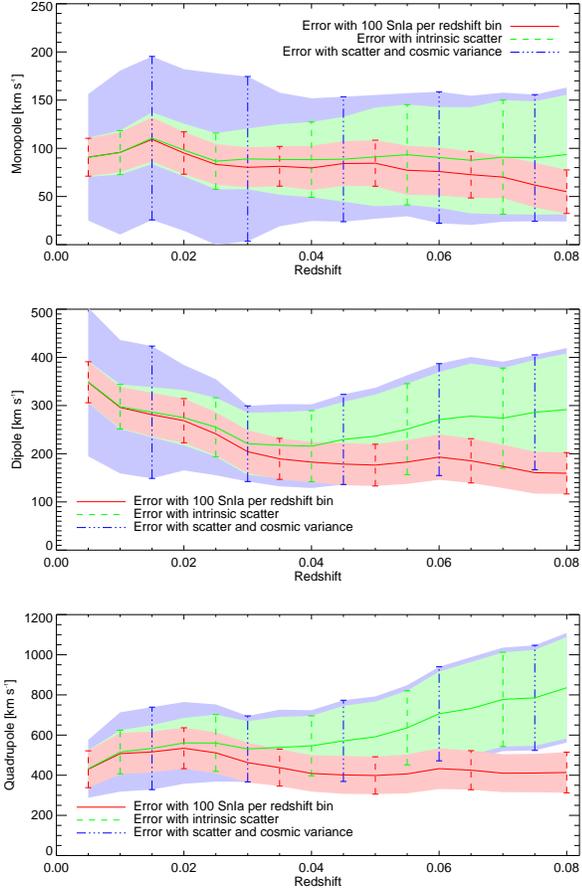}
\caption{Supernova rate proportional to density: The amplitudes of
the multipole vectors and the different errors in a SN Ia survey
with 100 SNe per redshift bin and an intrinsic scatter in the
magnitude of $\Delta m=0.08$.}\label{fig:multipoles2}
\end{center}
\end{figure}

We choose a set of $N=100$ SNe per bin in 16 redshift bins at redshifts of
$0.005-0.08$ or equivalently with Hubble flow velocities
$1500-24000 \kms$. This conforms roughly to the expectations from local
supernova searches conducted today \citep{li:2003,Krisciunas:2004,Jha:2006}
and in the near future \citep{aldering:2002,Frieman:2004,Hamuy:2006}, if they are
binned into 3 or 4 redshift bins. In
Figs.~\ref{fig:multipoles}-\ref{fig:multipoles2} we show the
evolution of the lowest multipoles as a function of $z$ for both
models of the supernova distribution. The red line and error bars
show how a hypothetical survey without any external error sources,
$\Delta m = 0$, would perform. Hence, the error here is only due to the
finite number of SNe that are used to probe the velocity field.
The green line shows the same, but including a Gaussian scatter of
$\Delta m = 0.08$ is included. Using the 27 different realisations
we can estimate the size of cosmic variance (shaded blue area).

From the figures it is clear that with 100 homogeneous distributed
SNe per redshift bin both the dipole and the quadrupole can be
measured out to a redshift of about 0.1. Furthermore for the synthetic observations,
we know the underlying cosmology and the real Hubble constant, and
we can determine the monopole. In the case of real observations we
can only measure the relative change. In other words the zero
point is in principle only measurable asymptotically at high
redshifts. Looking at the red curve, we see that for all
multipoles the power goes to zero at higher redshift, as the flow
approach the background Hubble flow. We also note that cosmic
variance is very large in the monopole term. This indicates that
the local ``Hubble Bubble'' phenomena, such as those found in
\citet{Zehavi:1998gz,jrk07}, are not unlikely.

The velocity amplitudes are positive definite, and including a
Gaussian scatter in the velocities, is adding uncorrelated
noise to the (synthetic) observations. Therefore the amplitudes of the
multipoles, when errors are included, are overestimated. This error
is per se hard to separate from the signal, but can trivially be
beaten down by a better control of the intrinsic errors or
increasing the number of SNe per redshift bin. Given an observational
data set synthetic observations with a realistic sky distribution
should be used to separate the noise amplitude from the underlying
velocity field.

In accordance with the underlying velocity distribution (Fig.~\ref{fig:sniahisto}),
on average there is a 5\% overestimation of the radial velocity amplitudes,
when assuming the Sn Ia rate is proportional to mass compared to a uniform
distribution.

\section{Results from the JRK sample of nearby supernovae}\label{sec:data}

We apply the analysis technique described in section \ref{sec:analysis} to the sample of 133 nearby
supernovae obtained by JRK. The JRK
sample includes 95 type Ia supernovae in the Hubble flow, with an
intrinsic dispersion of less than $7\%$ in distances. This is the
best sample available with distances derived in a homogeneous
way, by using the Multicolor Light Curve Shape (MLCS2k2) method
described in \citet{jrk07}.
This method is one of several others used to standardise
supernova distances, but is particularly powerful in that it
allows a disentangled correction for host galaxy extinction and
also provides a statistically reliable way to estimate the errors.

The entire JRK sample comprises 133 supernovae. We have selected 3
sub samples of these for our analysis. We followed \citet{jrk07}
in selecting  as the first sub sample 95 Hubble flow supernovae,
the {\it HF sample}, a selection based on distance cut,
requirement of an acceptable light curve fit and excluding objects
with very high extinction. The SNe in this sample have redshifts between $0.0085$
and $0.021$ and a weighted average redshift of $z=0.0196$ or $5900
\kms$. The second sub sample, the {\it 4500 sample}, includes 74
SNe, and is similar to the HF sample, but without the highest
redshift SNe and including a few SNe with a lower redshift. It has
an weighted average of $z=0.015$ or $4521 \kms$ and contains SNe
with redshift between 0.007 and 0.035. The last sub sample, the
{\it 3500 sample}, includes 42 SNe, and is similar to the 4500
sample, but without the highest redshift SNe. It has an weighted
average of $z=0.0118$ or $3550 \kms$ and contains SNe with
redshift between 0.007 and 0.017.

To the uncertainties in the distance moduli, we also add in
quadrature an additional error of 0.08 mag in order to properly
represent the final uncertainties following \citet{jrk07}.
In addition we add $50 \kms$ in quadrature to the errors in
the radial velocity in order to take the velocity dispersion
around the local anisotropic Hubble flow into account.
\citet{Karachentsev03} find that the radial velocity dispersion
around the local (anisotropic) Hubble flow within $5$ Mpc
amounts to only $41 \kms$, when distance errors are taken
into account.
For further details on the samples we refer to \citet{jrk07}.

\subsection{The effective window function}\label{sec:window}

\begin{figure}
\begin{center}
\includegraphics[width=0.4 \textwidth]{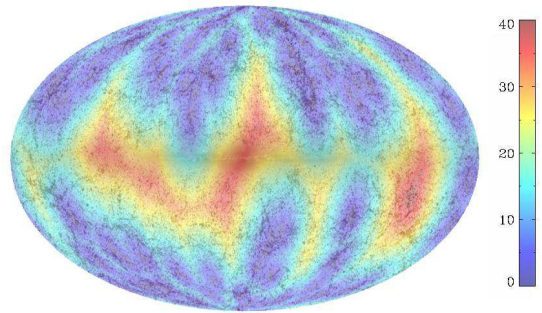}
\caption{The shortest distance on the sky to a supernova in
 degrees using the ``Hubble Flow'' sample}\label{fig:wf}
\end{center}
\end{figure}

In Fig.~\ref{fig:wf} we show the distance to the nearest supernova
on the sky, measured in degrees, for all points on the sphere.

The sample has good coverage, except for a few ``holes'' mainly
defined by the galactic disk. The mean distance to the nearest
supernova is $12.8^\circ$, but the maximum distance is
$40.5^\circ$ at $b=12.7^\circ$, $l=230.4^\circ$. There are three
areas with distance larger than $30^\circ$, roughly centred on
$(b,l)=(0^\circ,0^\circ),(5^\circ,80^\circ)$, and
$(10^\circ,230^\circ)$ respectively. For a given $l$, the distance
between zero points in the field is $180^\circ/l$. If the largest
holes in the sample have a size of $\Delta \theta$ then the
multipole decomposition becomes problematic around $l \sim
180^\circ/\Delta \theta$. For the present sample this corresponds
to $l \sim 2.5$ so that the quadrupole can be robustly fitted, but
not the octupole. We have tested this in practice. When the
quadrupole is added, the monopole and dipole amplitudes and the
dipole direction hardly change. However, when the octupole is
included there is serious leakage of power from the lower $l$'s to
$l=3$ and the results change substantially. The reason is that the
fit in the well-sampled regions can be improved by adding the
additional 7 $a_{3m}$ coefficients to the fit, but that this
happens at the expense of very large changes in the unsampled
regions. In order to probe the higher order multipoles it is
essential to reduce the size of the voids in the sample. For a
uniformly distributed sample of 95 supernovae the average distance
to the nearest supernova is 10.4 degrees, nearly the same as in
the JRK sample. However, the average maximum distance is 30.1
degrees, significantly lower than in our sample. A uniform sample
of 95 supernovae could be used to probe $l=3$ robustly. For a
uniform sample the average distance to the nearest supernova
scales as $N^{-1/2}$, and the average maximum distance in the
sample roughly as $N^{-0.4}$.

\subsection{Results}

\begin{figure}
\begin{center}
\includegraphics[width=0.4 \textwidth]{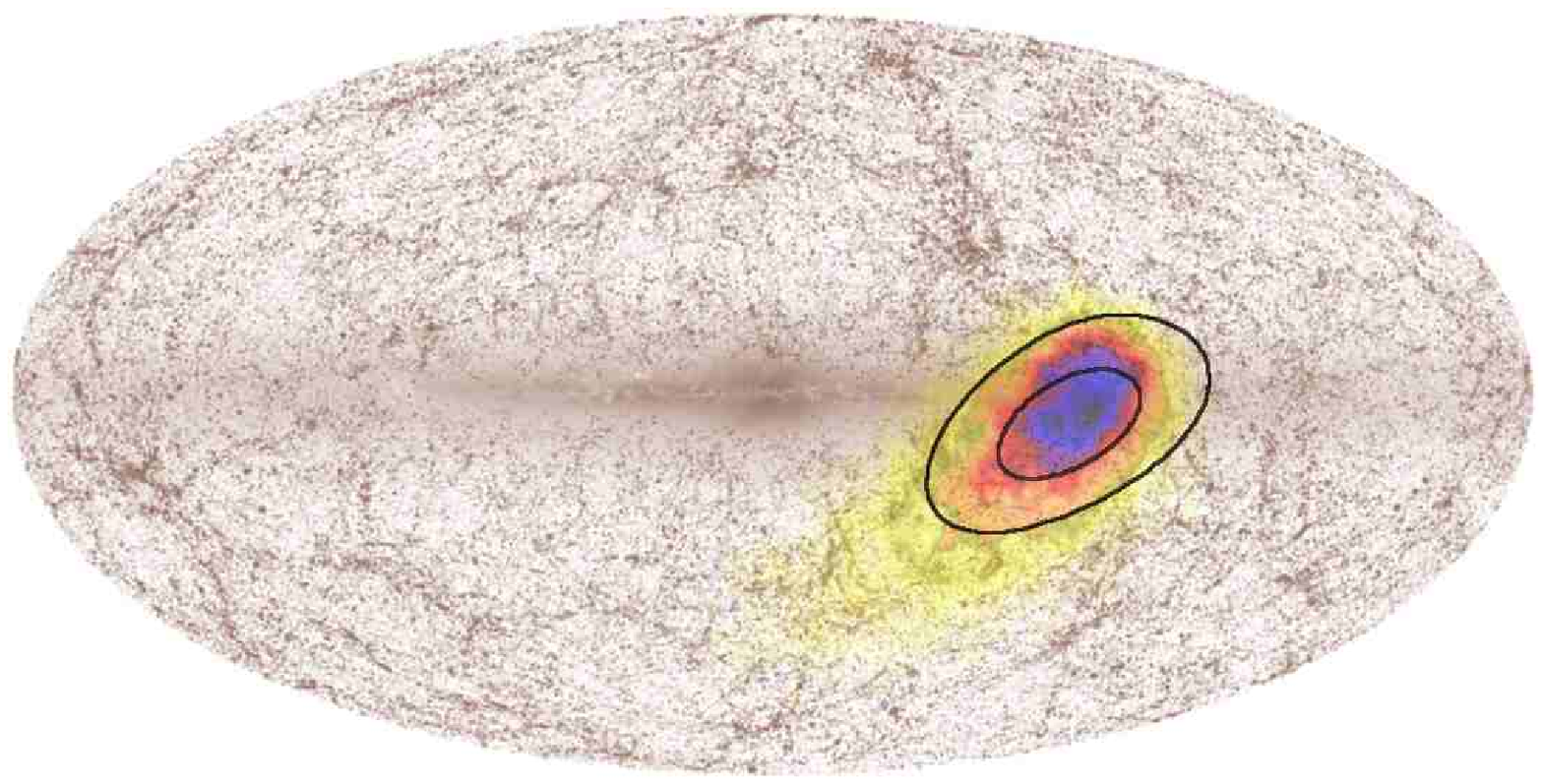}
\includegraphics[width=0.4 \textwidth]{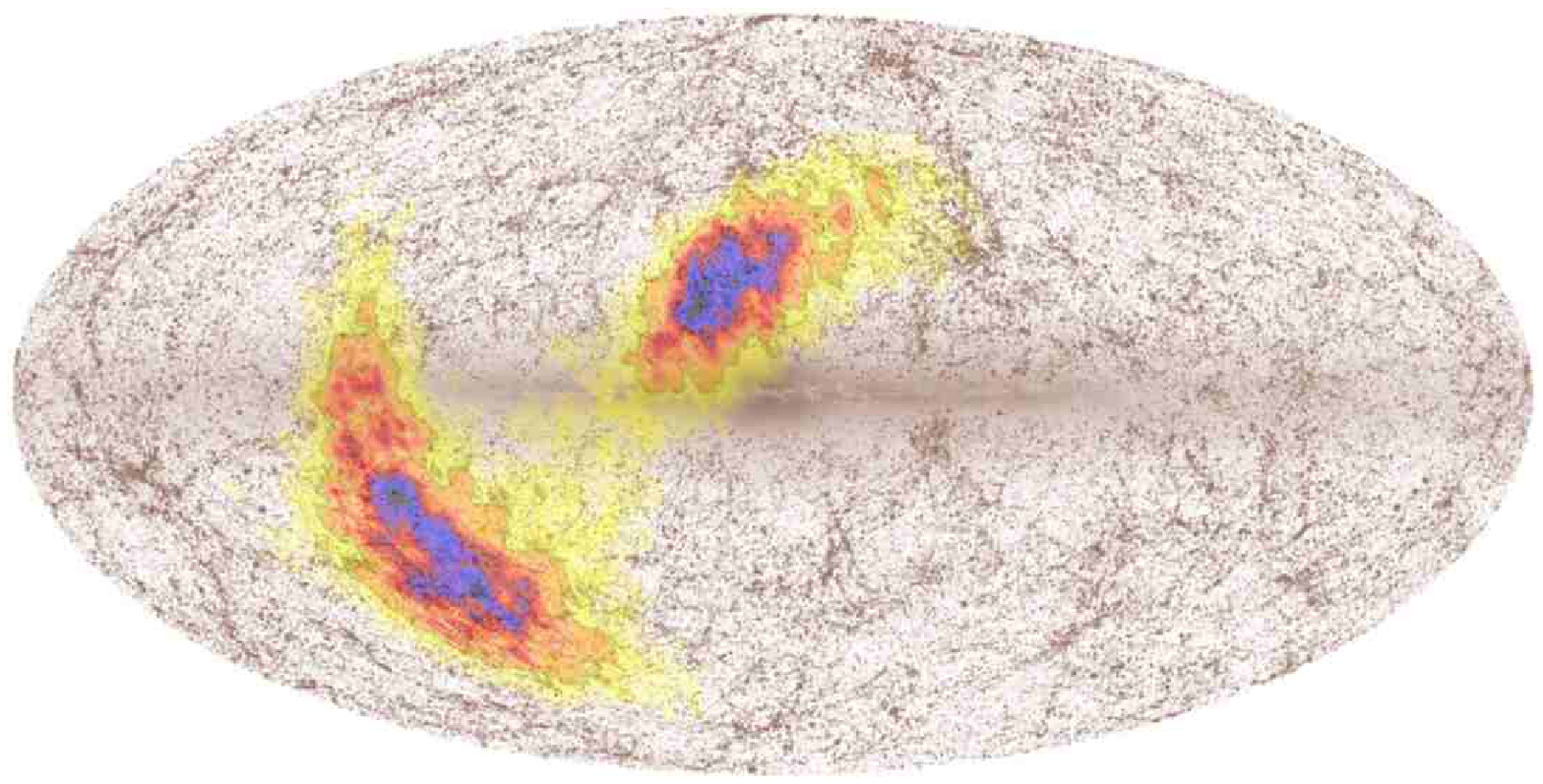}
\caption{The dipole vector (top) and polar quadrupole vector
(bottom) calculated from supernova data, with an weighted average
velocity of $3500 \kms$. The ellipses show the one
and two sigma errors}\label{fig:3500sample}
\end{center}
\end{figure}

\begin{figure}
\begin{center}
\includegraphics[width=0.4 \textwidth]{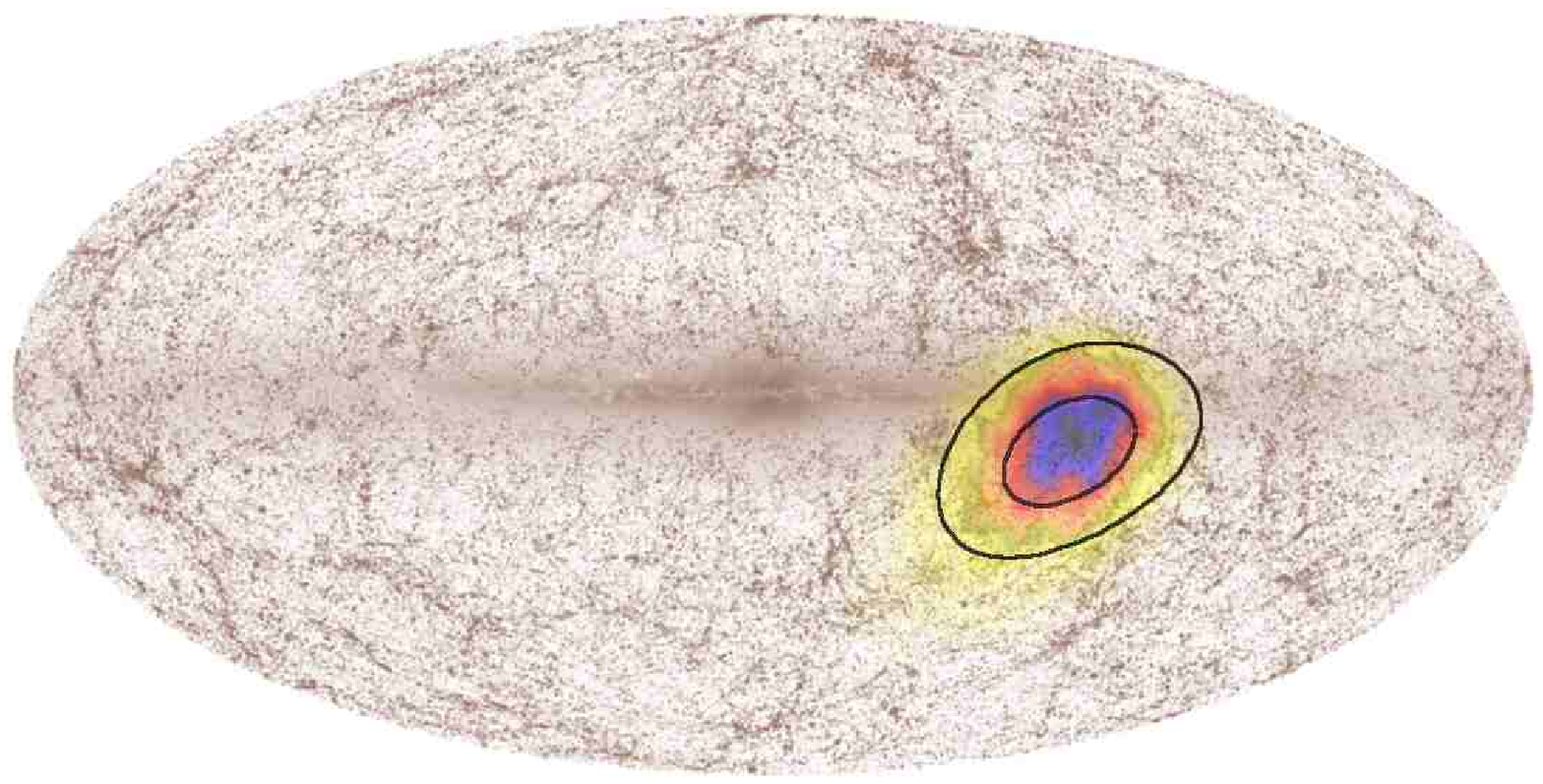}
\includegraphics[width=0.4 \textwidth]{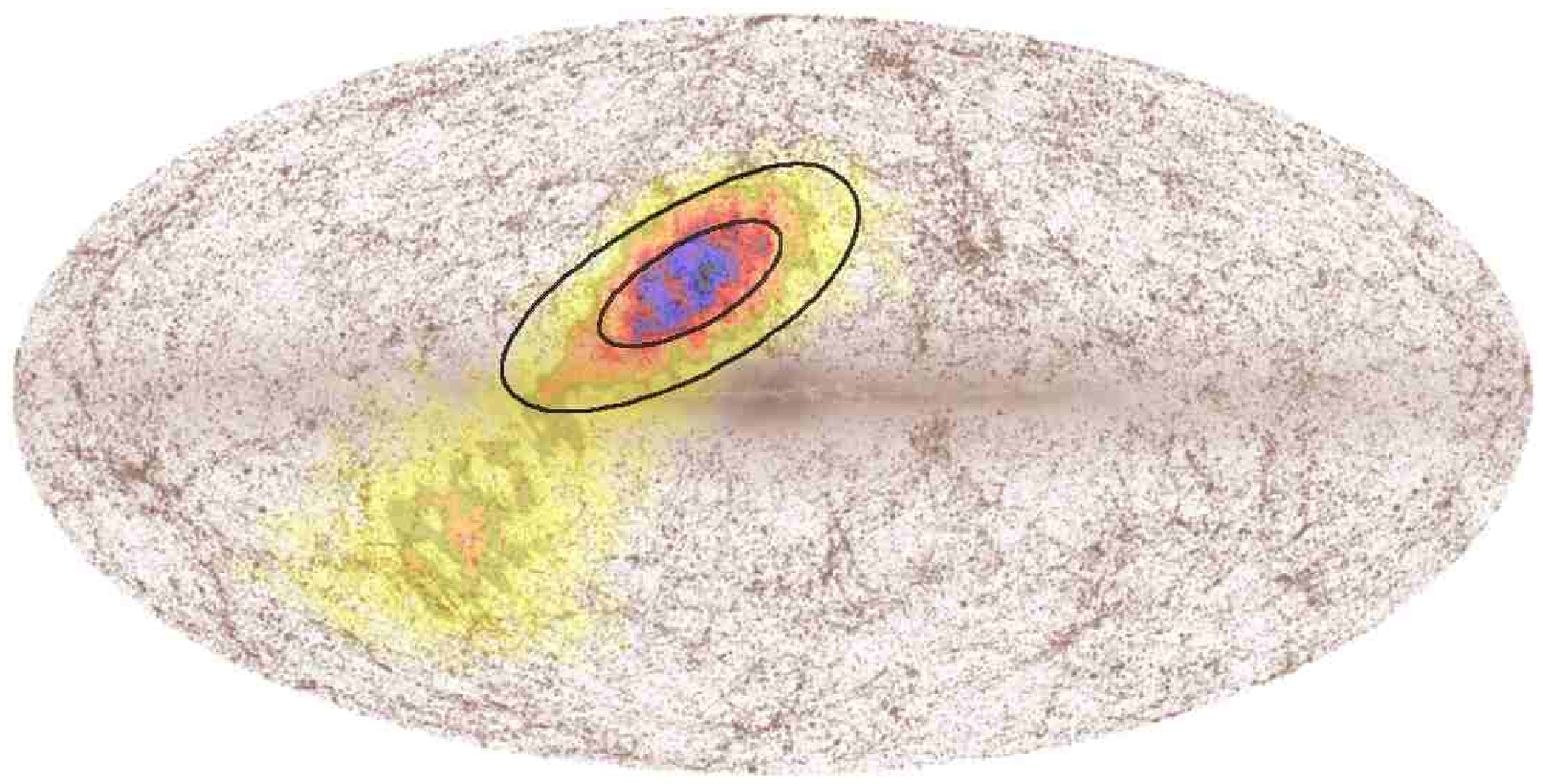}
\caption{The dipole vector (top) and polar quadrupole vector
(bottom) calculated from Sn Ia data, with an weighted average
velocity of $4500 \kms$. The ellipses show the one
and two sigma errors}\label{fig:4500sample}
\end{center}
\end{figure}

\begin{figure}
\begin{center}
\includegraphics[width=0.4 \textwidth]{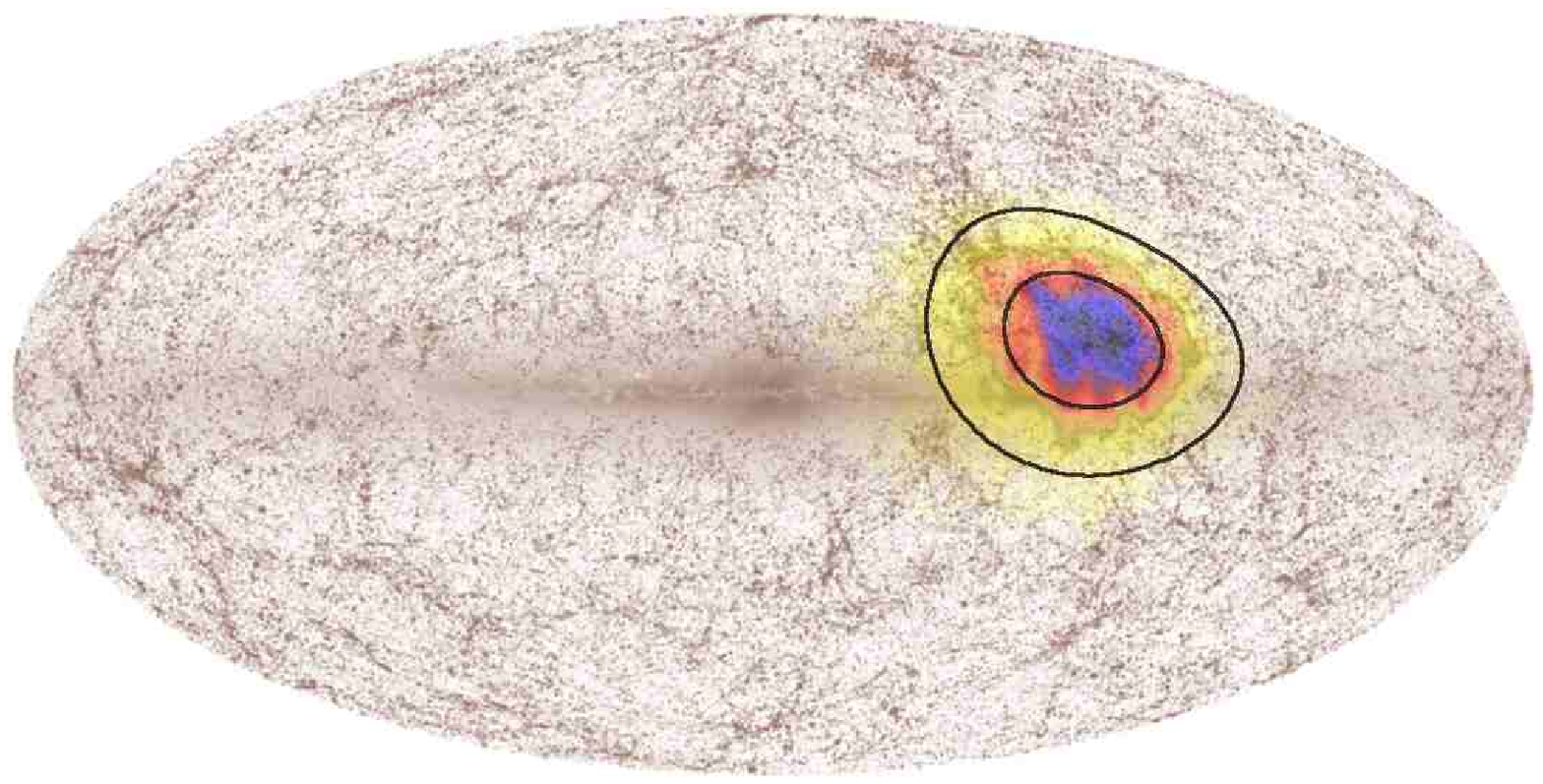}
\includegraphics[width=0.4 \textwidth]{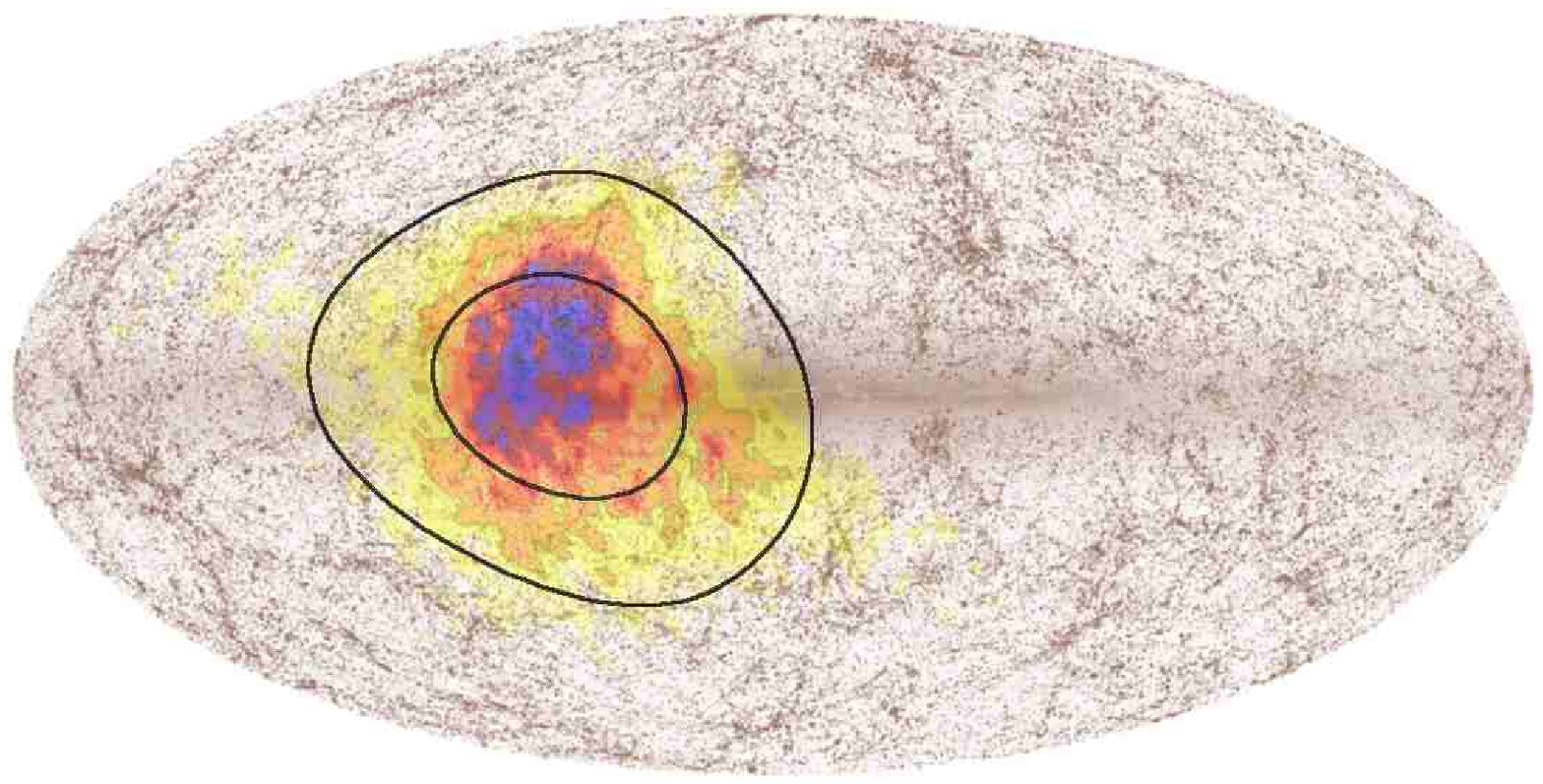}
\caption{The dipole vector (top) and polar quadrupole vector
(bottom) calculated from the JRK ``Hubble flow'' sample. The
ellipses show the one and two sigma errors}\label{fig:HFsample}
\end{center}
\end{figure}

In Figs.~\ref{fig:3500sample}-\ref{fig:HFsample} we show the
obtained 68\% and 95\% contours for the direction of the dipole
and quadrupole vectors in galactic coordinates. The corresponding
best fit values with their formal 68\% errors are shown in Tables
\ref{table:dipole} and \ref{table:quadrupole}.

\begin{figure}
\begin{center}
\includegraphics[width=0.4 \textwidth]{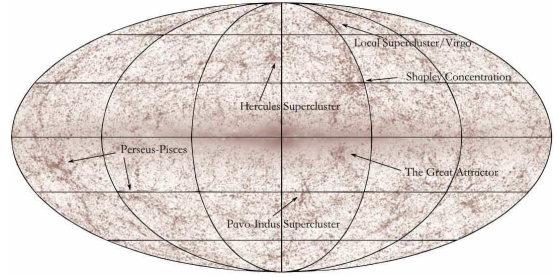}
\caption{The local universe as seen by the 2MASS survey. The
galaxy distribution image is courtesy of Dr. T.H. Jarrett
(IPAC/Caltech) and the 2MASS team.  Complete image can be found
at: http://spider.ipac.caltech.edu/staff/jarrett/papers/LSS. The
arrows indicate important super clusters.}\label{fig:LSS}
\end{center}
\end{figure}

\begin{figure}
\begin{center}
\includegraphics[width=0.4 \textwidth]{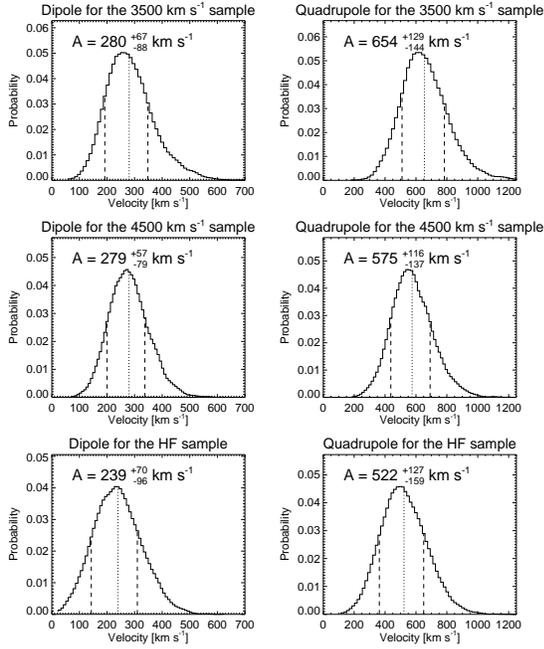}
\caption{The one dimensional distributions of the velocity amplitudes for the
different samples, with the 68\% limits indicated by a dashed line and
the median value by a dotted line.}\label{fig:ad}
\end{center}
\end{figure}

\begin{figure}
\begin{center}
\includegraphics[width=0.4 \textwidth]{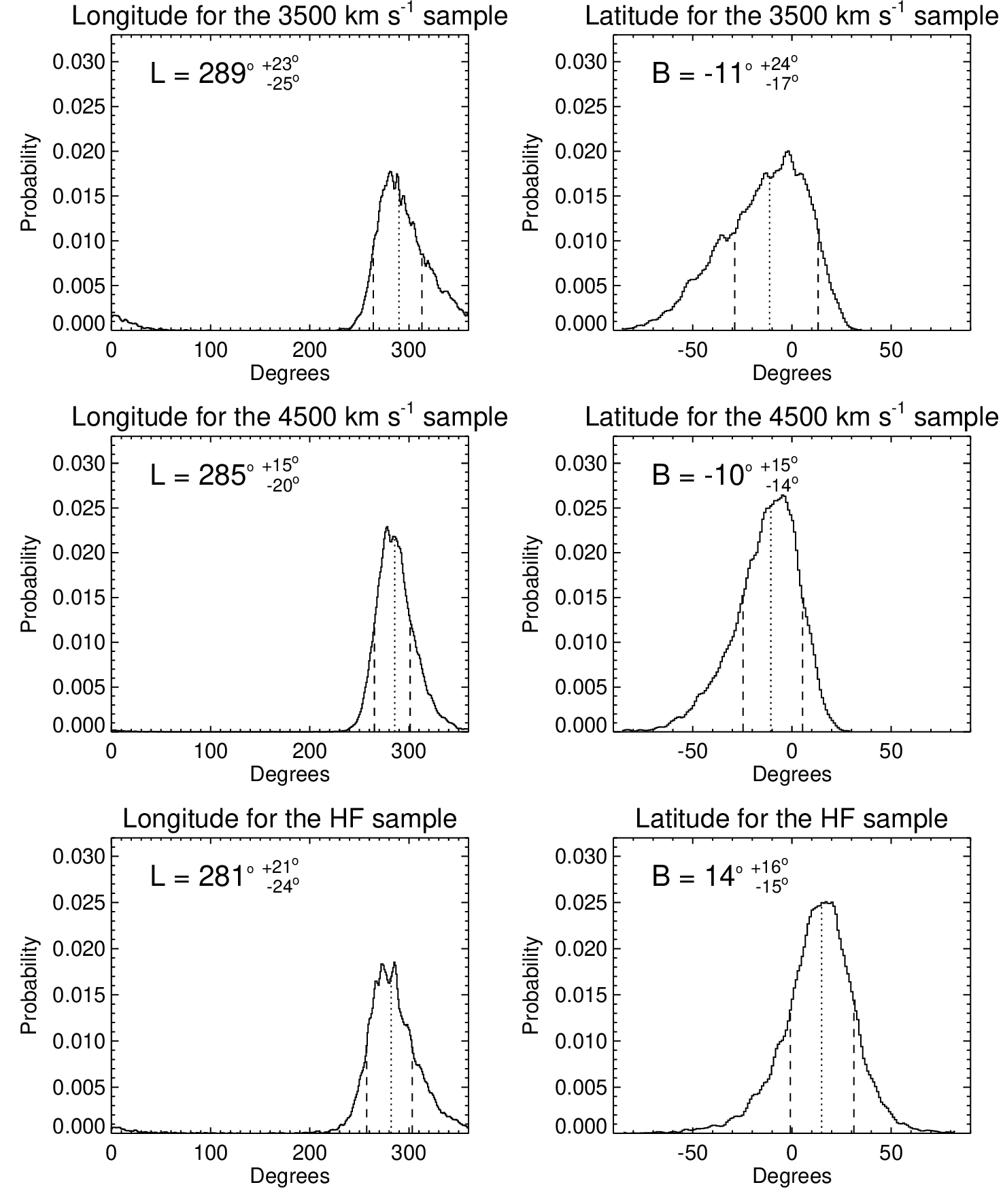}
\caption{The one dimensional distributions of the dipole vectors
for the different samples, with the 68\% limits indicated by a
dashed line and the median value by a dotted line.}\label{fig:dpd}
\end{center}
\end{figure}

\begin{figure}
\begin{center}
\includegraphics[width=0.4 \textwidth]{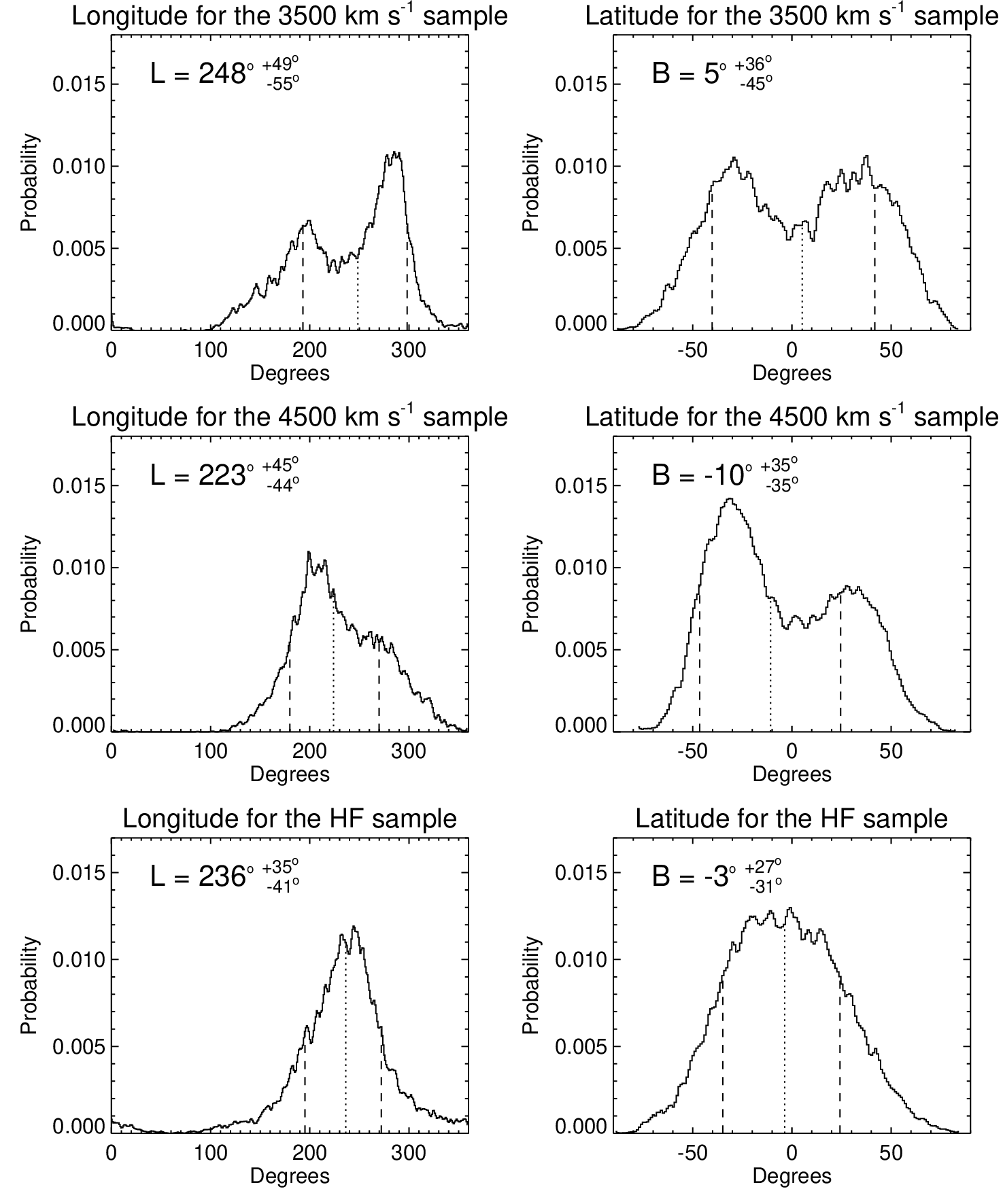}
\caption{The one dimensional distribution of the polar quadrupole
vectors for the different samples, with the 68\% limits indicated
by a dashed line and the median value by a dotted line.}\label{fig:qpd}
\end{center}
\end{figure}

This result can be compared to other velocity surveys based on
galaxy samples. The recent review by \citet{feldman} summarises
these surveys. At an effective depth of 4000 $\kms$ they find that
the dipole amplitude is $330 \pm 101 \kms$ in direction $l =
234^\circ \pm 11^\circ$ and $b=12^\circ \pm 9^\circ$.

\subsubsection{Dipole}

The 3500 and 4500 samples both show a dipole in a direction
compatible with the Great Attractor region at $l \sim 300^\circ$,
$b \sim 0^\circ$. For the HF sample the direction shifts to
slightly higher $b$, compatible with a shift in the motion towards
the Shapley concentration \citep[which lies at an average
distance of 14000 $\kms$,][]{Bardelli} at slightly higher galactic latitude.
Furthermore the amplitude of the dipole decreases, as is expected
from the Monte Carlo simulations.

\subsubsection{Quadrupole}

For all the three sub samples we find a relatively large
contribution from the quadrupole, showing that the local flow has
a significant shear component. The result is consistent in
magnitude with the expectation from the Monte Carlo simulations.
From the figures it can also be seen that there is a change in
quadrupole direction with redshift, and that the distributions for
the 3500 and 4500 samples are bimodal. This could be because
the quadrupole is pointing in different directions at the lower and higher
end of the included redshift range of the SNe, and hence for the 3500
sample the bimodality is more apparent. Even though we quote formal
68\% errors in Table 2 the distribution is highly non-Gaussian and
the error bars should be taken as indicative only.

\begin{table}[h]
\begin{center}
\begin{tabular}{cccc}
\hline \hline Sample & $v_r [\kms]$ & $l$ & $b$ \cr
\hline 3500 & \WE{280}{+67}{-88} &
\WE{289^\circ}{+23^\circ}{-25^\circ} &
\WE{-11^\circ}{+24^\circ}{-17^\circ} \cr 4500 & \WE{279}{+57}{-79}
& \WE{285^\circ}{+15^\circ}{-20^\circ} &
\WE{-10^\circ}{+15^\circ}{-14^\circ} \cr HF & \WE{239}{+70}{-96} &
\WE{281^\circ}{+21^\circ}{-24^\circ} &
$\phantom{-}$\WE{14^\circ}{+16^\circ}{-15^\circ} \cr \hline \hline
\end{tabular}
\end{center}
\caption{Amplitude and direction of the dipole vector for the three different
samples.} \label{table:dipole}
\end{table}

\begin{table}[h]
\begin{center}
\begin{tabular}{cccc}
\hline \hline Sample & $v_r [\kms]$ & $l$ & $b$ \cr
\hline 3500 & \WE{654}{+129}{-144} &
\WE{248^\circ}{+49^\circ}{-55^\circ} &
\WE{5^\circ}{+36^\circ}{-45^\circ} \cr
       4500 & \WE{575}{+116}{-137} & \WE{223^\circ}{+45^\circ}{-44^\circ} & \WE{-10^\circ}{+35^\circ}{-35^\circ} \cr
         HF & \WE{522}{+127}{-159} & \WE{236^\circ}{+35^\circ}{-41^\circ} & \WE{-3^\circ}{+27^\circ}{-31^\circ} \cr  \hline \hline
\end{tabular}
\end{center}
\caption{Amplitude of the quadrupole, and direction of the polar quadrupole vector for the three
different samples.}\label{table:quadrupole}
\end{table}

\section{Comparison with other results}

\subsection{SNIa}

The \citet{tonry:2003} data set of 98 SNe was analysed by
\citet{hudson} in order to find the local dipole. For the
part of the sample with $v_r < 6000 \kms$ they found a dipole of
$v_r = 376 \pm 81\kms$ towards $l = 285^\circ$,
$b=-14^\circ$. They do not quote error bars on this result, but
for the part of the sample with $v_r > 6000 \kms$ the
stated errors are $\pm 17^\circ$ for $l$ and $\pm 13^\circ$ for
$b$. This result is completely compatible with our result for the
dipole. However, Hudson et al.~quote no results for the higher
order terms.

\citet{jrk07} also provide a crude estimate of the dipole
amplitude and direction from a subset of 69 supernovae in their
sample which is roughly compatible with our 4500 sample . In the
coordinate system of the Local Group they find a velocity of
$541\pm75 \kms$ towards a direction of $(l,b) = (258^\circ
\pm 18^\circ,51^\circ \pm 12^\circ)$. If we transform our dipole
term from the HF sample to the same coordinate system, using the
Local Group velocity derived in \cite{rauzy}, we find a velocity
of $516 \kms$ towards $(b,l) = (248^\circ,51^\circ)$. Both
the amplitude and direction are compatible with the JRK value at
1$\sigma$. Our derived amplitude is slightly lower (although not
significantly so), because in our fit the quadrupole term accounts
for part of the velocity.

As noted before, we do not discuss the local monopole term. By
subdividing supernovae into low and high redshift bins
\citep{Zehavi:1998gz,jrk07}, a significant variation in the local
Hubble parameter has been detected. This does not have any impact
on our results (see section \ref{sec:window}), and since it was discussed thoroughly in
\citet{jrk07} we refer to that paper for further details.

\subsection{Galaxy surveys}

Results from galaxy velocity surveys on scales of order 4000-6000
$\kms$ generally agree that the magnitude of the dipole is of
order 300 $\kms$ in the direction $l \sim 300^\circ$, $b \sim
20^\circ$ (see for instance \citet{zaroubi} and references
therein). This result is compatible with the SN Ia dipole
direction and magnitude within $2\sigma$.

A reconstruction of the very local velocity field ($<3000 \kms$)
was done by \citet{tonry:2000} measuring surface brightness
fluctuations in 300 early type galaxies, predominantly in groups and
clusters. They used an explicit flow model with a Virgo Attractor
and a Great Attractor which contain the main local mass
concentrations. Furthermore they added dipole and quadrupole terms
to account for the gravitational pull and shear from large scale
structure further away. They find a very low value of the dipole
($\sim 150 \kms$) and the quadrupole polar vector ($\sim 50
\kms$), but it may be related to having the dipole in the same
direction as the attractors, and the attractors accounting for the
major part of the shear (quadrupole term) in the model.

The dipole has also been measured using velocity field
reconstruction of the 2mass catalogue. At a distance of 4000-6000
$\kms$ the dipole direction is found to be roughly $l \sim
250^\circ$, $b \sim 35-40^\circ$, again compatible with our result
within $2\sigma$ \citep{Erdogdu:2006nd,Erdogdu:2006qs,pike}.

\subsection{Clusters}

Cluster samples like SMAC probe larger distances, and find
directions which are generally compatible with the SN Ia result.
For example \citet{hudson} find $l = 260^\circ \pm 13^\circ$,
$b = 0^\circ \pm 11^\circ$. However, they find an amplitude of
$687 \pm 203 \kms$, significantly higher than our result (although
again compatible at $2\sigma$).

\section{Discussion}

We have analysed mock supernova surveys in order to study the
number of supernovae needed to probe the large scale velocity
field of the local universe, quantified in terms of the angular
power spectra as a function of redshift. We then proceeded to use
the best available database of low-redshift supernovae, the JRK
sample, to probe the local dipole and quadrupole of
the velocity field at three different distances.
The present method has several advantages over galaxy
surveys. The uncertainty on each individual supernova luminosity
is much smaller than the systematic uncertainties in determining
galaxy luminosities so that a much smaller sample is sufficient.
We find that
\begin{itemize}
\item{With two different models for the type Ia supernova rate, the resulting mock surveys
only differ at the percent level. Hence, using SN Ia to probe the underlying
velocity field is robust with respect to assumptions about the supernova environment.}
\item{For the dipole we find a result which is consistent with
galaxy surveys at the same Hubble flow depths.}
\item{The quadrupole is comparable in value to the dipole, indicative of a
significant shear in the local velocity field, in accordance with our mock catalogues.
It has, to our knowledge, not been measured before at these distances.}
\item{With the present sample size of almost 100 supernovae the precision
of the dipole measurement is comparable to that in galaxy surveys
using thousands of galaxies.}
\end{itemize}

Finally, we note that new surveys like Pan-STARRs, SkyMapper and LSST will
measure about 10,000 type Ia supernovae at $z < 0.1$ per year, and
if proper light curves and redshifts can be measured for even a small fraction
of these events they will provide an extremely powerful tool for
studying the dynamics of the local universe.

\section*{Acknowledgements}
We thank the Danish Centre of Scientific Computing (DCSC) for granting
the computer resources used. TH thanks the DARK Cosmology Centre for
hospitality during the course of this work. SJ is grateful for support at KIPAC
and SLAC via the Panofsky Fellowship. The Dark Cosmology Centre is funded
by the DNRF.


\begin{thebibliography}{99}
\bibitem[{{Aldering et al.}(2002)}]{aldering:2002} Aldering, G., et al.\
2002, \procspie, 4836, 61

\bibitem[{{Bardelli et~al.~}(1994)}]{Bardelli}
Bardelli, S., et al.\ 1994, \mnras, 267, 665

\bibitem[{{Bonvin et al.}(2006a)}]{Bonvin:2005ps} Bonvin, C., Durrer, R.,
\& Gasparini, M.~A.\ 2006, \prd, 73, 023523

\bibitem[{{Bonvin et al.}(2006b)}]{Bonvin:2006en} Bonvin, C., Durrer, R.,
\& Kunz, M.\ 2006, \prl, 96, 191302

\bibitem[{{Cooray \& Caldwell}(2006)}]{Cooray:2006ft} Cooray, A., \&
Caldwell, R.~R.\ 2006, \prd, 73, 103002

\bibitem[{{Copi et al.}(2006){Copi, Huterer, Schwarz, and Starkman}}]{copi06}
Copi, C.~J., Huterer, D., Schwarz, D.~J., \& Starkman, G.~D.\ 2006,
\mnras, 367, 79

\bibitem[{{Erdo{\u g}du et al.}(2006a)}]{Erdogdu:2006nd} Erdo{\u g}du, P.,
et al.\ 2006, \mnras, 373, 45

\bibitem[{{Erdo{\u g}du et al.}(2006b)}]{Erdogdu:2006qs} Erdo{\u g}du, P.,
et al.\ 2006, \mnras, 368, 1515

\bibitem[{{Feldman, Sarkar \& Watkins} (2006)}]{feldman}Sarkar, D., Feldman, H.~A. \& Watkins,
R. 2006, astro-ph/0607426

\bibitem[Frieman et al.(2004)]{Frieman:2004} Frieman, J., et al.\ 
2004, BAAS, 36, 1548

\bibitem[Hamuy et al.(2006)]{Hamuy:2006} Hamuy, M., et al.\ 2006, 
\pasp, 118, 2

\bibitem[{{Hudson et al.}(2004)}]{hudson}Hudson, M.~J., Smith, R.~J.,
Lucey, J.~R., and Branchini, E. 2004, \mnras, 352, 61

\bibitem[{{Hui \& Greene}(2006)}]{Hui:2005nm} Hui, L., \& Greene,
P.~B.\ 2006, \prd, 73, 123526

\bibitem[Jha et al.(2006)]{Jha:2006} Jha, S., et al.\ 2006, \aj,
131, 527

\bibitem[{{Jha et al.}(2007){Jha, Riess \& Kirshner}}]{jrk07}
Jha, S., Riess, A.~G.~, and Kirshner, R.~P.\ 2007, submitted to \apj

\bibitem[{Karachentsev et al.}(2003)]{Karachentsev03}
Karachentsev, I.~D., Makarov, D.~I., Sharina, M.~E., et al.\ 2003, \aap, 398, 479

\bibitem[Krisciunas et al.(2004)]{Krisciunas:2004} Krisciunas, K., et
al.\ 2004, \aj, 128, 3034

\bibitem[Li et al.(2003)]{li:2003} Li, W., et al.\ 2003, \pasp,
115, 453

\bibitem[{{Miller \& Branch}(1992)}]{miller:1992}
Miller, D.~L. and Branch, D.\ 1992, \aj, 103, 379

\bibitem[{{Neill et al.}(2006)}]{Neill:2006} Neill, J.~D., et al.\
2006, \aj, 132, 1126

\bibitem[{{Pike \& Hudson}(2005)}]{pike}Pike, R.W. \& Hudson,
M.~J.\ 2005, \apj, 635, 11

\bibitem[{{Rauzy \& Gurzadyan}(1998)}]{rauzy}Rauzy, S. \& Gurzadyan,
V.~G. 1998, \mnras, 298, 114

\bibitem[{{Riess et al.}(1995)}]{Riess:1995} Riess, A.~G., Press,
W.~H., \& Kirshner, R.~P.\ 1995, \apjl, 445, L91

\bibitem[{{Riess et al.}(2004)}]{riess:2004}
Riess, A.~G., Strolger, L.-G., Tonry, J. et al.\ 2004, \apj, 607, 665

\bibitem[{{Sarkar, Feldman \& Watkins}(2006)}]{Sarkar:2006gh}
  Sarkar, D., Feldman, H.~A. and Watkins, R.\ 2006,
  arXiv:astro-ph/0607426

\bibitem[{{Sharon et al.}(2006)}]{sharon}Sharon, K.\ et al.\ 2006, astro-ph/0610228

\bibitem[{{Shi \& Turner}(1997)}]{Shi:1997aa}
  Shi, X.~D. and Turner, M.~S.\ 1998,
  \apj, 493, 519

\bibitem[{{Springel}(2005)}]{Springel:2005mi}
  Springel, V.\ 2005,
  \mnras,  364 1105

\bibitem[{{Springel, Yoshida \& White}(2000)}]{Springel:2000yr}
  Springel, V., Yoshida, N. and White, S.~D.~M.\ 2001,
  New Astron.,  6, 79

\bibitem[{{Sugiura, Sugiyama \& Sasaki}(1999)}]{sugiura:1999}
Sugiura, N., Sugiyama, N., \&  Sasaki, M.\ 1999,
  Prog.\ of Theoretical Physics, 101, 903

\bibitem[{{Sullivan et al.}(2006)}]{sullivan} Sullivan, M., et al.\
2006, \apj, 648, 868

\bibitem[Tonry et al.(2000)]{tonry:2000} Tonry, J.~L., Blakeslee,
J.~P., Ajhar, E.~A., \& Dressler, A.\ 2000, \apj, 530, 625

\bibitem[{{Tonry et al.}(2003)}]{tonry:2003}Tonry, J.~L.\ et al.\ 2003, \apj, 594, 1

\bibitem[{{Zaroubi}(2002)}]{zaroubi}Zaroubi, S.\ 2002,
arXiv:astro-ph/0206052

\bibitem[{{Zehavi et al.}(1998)}]{Zehavi:1998gz}
  Zehavi, I., Riess, A.~G., Kirshner, R.~P. and Dekel, A.\ 1998,
  \apj, 503, 483

\end{thebibliography}
\end{document}